# An Exploratory Study of Health Habit Formation Through Gamification

**Anna Iurchenko**
*Product/UX Designer at Stanfy, San Francisco, USA*
aiurchenko@stanfy.com

## ABSTRACT

Promotion of health habits help maintain and improve people's health, reduce disease risks, and manage chronic illness. Regular healthy activities like walking, exercising, healthy eating, drinking water or taking medication on time require forming the new habits.

Gamification techniques are promising in promoting healthy behaviors and delivering health promotion information. However, using 'gaming' elements such as badges, leader boards, health-related challenges in mobile applications to motivate and engage people to change health behavior is quite new. In this exploratory study we aimed to assess how game mechanics and dynamics influence formation of a habit through the mobile application.

## INTRODUCTION

Gamification means the addition of game elements (such as badges, leader boards, challenges, rewards, the ability to 'level up' and use of avatars) to non-game contexts [1, 2,] as a means to motivate and increase user activity and retention. Gamification approach in the health-related mobile applications has a potential of changing people's behavior and influencing forming of the new habits [3, 4].

Games are designed to motivate users behavior and learnings from behavioral economics are related to many of these features [5, 6]. For example, many games provide conditional rewards (e.g. points and prizes) that risk being lost if gamers do not frequently return to play. This plays on the well-known tendencies of people to avoid losses (loss aversion) and to irrationally value things they hold over things they do not have (endowment effect).

Several health-related behaviours have been shown to have a habitual component, including hand hygiene [17], medication adherence [19], and brushing teeth [17, 18]. Mobile phones have been proven to be effective for helping to build the health-related habits, like for smoking cessation and encouraging medication adherence [7, 8, 9].

Habits develop through context-dependent repetition: repeated action in a particular context reinforces association, such that alternative responses become less accessible and the habitual response proceeds automatically upon encountering relevant cues [20]. Automaticity is the defining feature of habit: unlike intentional action, which often requires conscious effort, habit is characterized by direct environmental cuing of behavior [21]. Consequently, where habits conflict with deliberative intentions, behaviour tends to be guided by habit and not intention [23]. Habit formation may thereby sustain new behaviours when motivation is lost [22].

Some of the best examples of gamification are games that encourage exercise by turning physical activity into a game [10]. Also the use of mobile phones for these games shown advantages that mobile apps can create to support and monitor outdoor activities.

The aim of this exploratory study was to investigate the potential role of gamification methods on behavioural and habit formation, based on analyzing data of participants instructed to drink water daily to maintain body hydration [11, 12] and reporting it through the mobile application.

In this study we aimed to remind participants to drink water regularly throughout the day with the help of the mobile application. We aimed to support repetition in a manner conducive to automaticity. We manipulated the gamification elements to investigate its impact on behaviour and final automaticity [24].

## METHOD

To apply gamification it is important to understand which game design elements will have the biggest impact on the behavior and then define the game flow and integrate it into the usage scenarios.

The technology works when it employs specific behavior change ingredients, as one of the key principles of evidence based behavioral medicine [13, 14]. These persuasive ingredients should encourage them to shift their beliefs, attitudes, and actions.

There are 7 core ingredients of gamification that have clear linkages to proven behavior change strategies, with the exception of fun and playfulness, which has perhaps, not received much attention in the health behavior change literature [15].

The persuasive architecture of gamification is the combination of ingredients that make a product fun and engaging. In Table 1 the popular gamification mechanics are listed.

One of the well-accepted theory is that players in any experience are seeking mastery. In the original research by Dreyfus for the U.S. Army (revised in 1990) [16], a series of stages of mastery emerged when looking at how people engage with the system. Gamification strategy for this study was designed around core levels outlined by Dreyfus.

In this study, we adopted this theory to influence health-related outcomes and designed the strategy as it shown in Figure 1 and Table 2

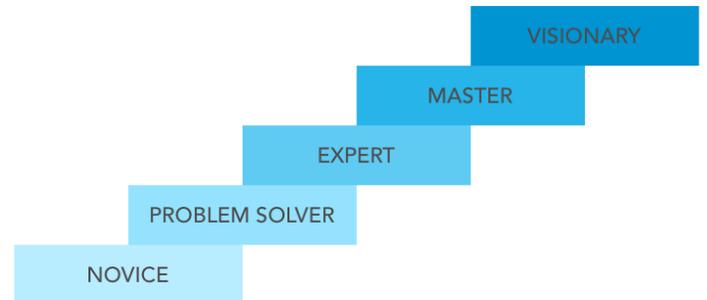

FIGURE 1. Mastery of a system. Rising from novice to visionary across a series of steps.

No participant should be obliged to expected to progress to visionary - the system should enable the player to stick to the habit as soon as possible.

For gamification to be effective, gamified technology must outperform other design patterns, regarding its ability to influence people's beliefs, attitudes, or behaviors. Moreover, gamification must sustain these impacts over the long-term, and offer more than a short-term novelty effect.

Participants were challenged to "fill in" the body on the picture (the silhouette was changed depending on user's gender) to drink eight glasses of water every day to level up. If they didn't drink eight glasses, they stayed at the same level.

| Game element | Persuasive strategy | Gamification tactics (on-screen features that users interact with) |
|---|---|---|
| Goal setting | Committing to achieve a goal | Providing clear goals |
| Capacity to overcome challenges | Growth, learning, and development | Offering a challenge, Using levels (incremental challenges) |
| Providing feedback on performance | Receiving constant feedback through the experience | Providing feedback, Allocating points |
| Reinforcement | Gaining rewards, avoiding punishments | Giving rewards, Providing badges for achievements |
| Compare progress | Monitoring progress with self and others | Showing progress |
| Social connectivity | Interacting with other people | Showing the game leaders |
| Fun and playfulness | Paying out an alternative reality | Giving a story or theme |

TABLE 1. The persuasive architecture of gamification and popular gamification tactics.

| Levels | Activities |
|---|---|
| Novice | Fill in profile, add day schedule so we could prompt about water intake at the right time. Checked all tips about health benefits of drinking watere |
| Problem solver | Maintain water balance for 1 day, 3 days, 7 days |
| Expert | Maintain water balance for 14 days in a row |
| Master | Maintain water balance for 21 days in a row |
| Visionary | Kept water balance rate for 3 months |

TABLE 2. Gamification levels and corresponding users activities.

Fifty participants were recruited for this study. Twenty-five participants were instructed to use application that allowed them to track their water intake and notified throughout the day to drink a glass of water.

Another twenty-five were using the same mobile application with the same functionality but with gamification elements added.

All participants completed the study up to and including 4-week follow-up. Usage data was gathered through the database and participants were surveyed every week.

## RESULTS

Results indicate different levels of user engagement depending on whether participants used the app with the gamification elements (group A) or without them (group B) (Figure 2).

This usage data captured from all the participants on the server. Twenty-five percent (25%) of participants from the group A were maintaining their water balance after four weeks, and only around ten percent (10%) showed the same results from the group B.

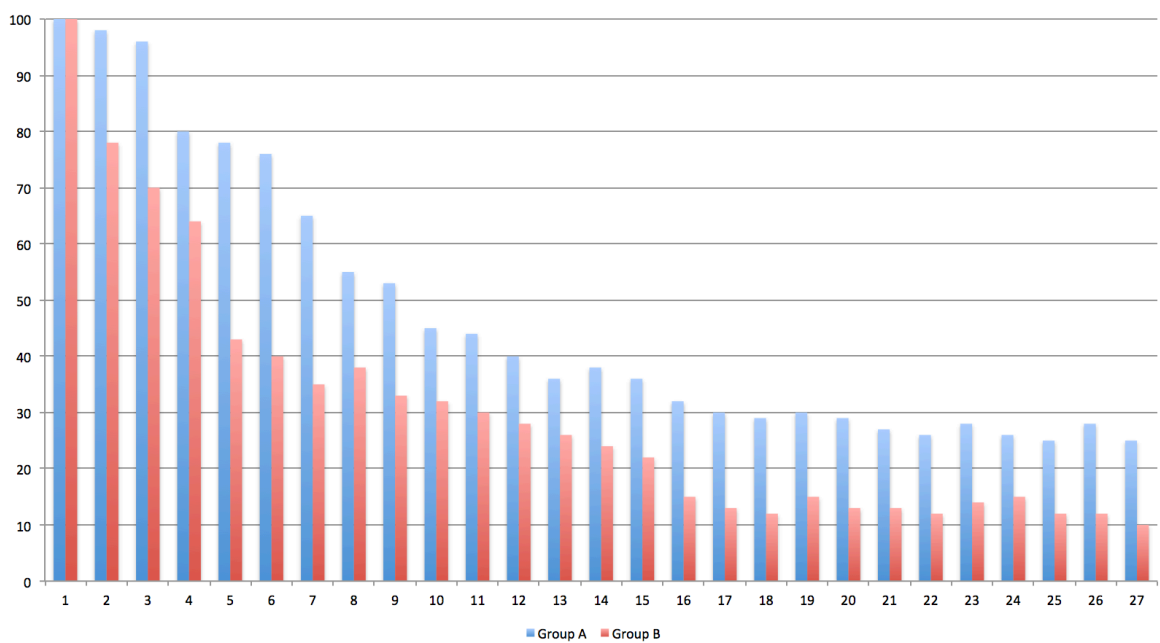

FIGURE 2. The rate of participants who maintain the water balance each day of the study, separated by group (Group A - used the app with gamification elements, Group B - used app without gamification elements).

These results also indicate three types of users: those who tried the application, those who engaged for at least one week and those who chose to engage with the application for longer than a week. For the last type, the results were higher for the Group A by 10%.

**DISCUSSION AND CONCLUSIONS**

This exploratory study demonstrates that there are some promising links between gamification principles and health behavior change. Results indicate the different level of user engagement depending on the presence of gamification elements and suggest that there is value in adding game elements to the user experience. However, at this stage of the research, we could not tell if the reported outcomes represent sustainable long-term impacts or just short-term effects.

It is easy to see how existing digital interventions can borrow gamification principles, by considering flow, meaningful rewards, making them more social, and most importantly, finding innovative ways to make health habits forming fun and engaging.